\documentclass[10pt,10pt, journal, onecolumn]{IEEEtran}
\usepackage[latin1]{inputenc}
\usepackage{array}
\usepackage{mathtools}
\usepackage{multirow}
\usepackage{amsmath}
\usepackage{amsthm}
\usepackage{amssymb}
\usepackage{graphicx}
\usepackage{esint}
\PassOptionsToPackage{normalem}{ulem}
\usepackage{ulem}

\makeatletter

\providecommand{\tabularnewline}{\\}

\theoremstyle{plain}
\newtheorem{thm}{\protect\theoremname}
\theoremstyle{remark}
\newtheorem{rem}[thm]{\protect\remarkname}

\usepackage{algorithm}
\usepackage{algorithmic}
\usepackage{epsfig}
\usepackage{cite}
\usepackage{array}
\usepackage{fancyhdr}
\usepackage{epsf}
\usepackage{epsfig}
\usepackage{rotating}
\usepackage{longtable}
\usepackage{bm}
\usepackage{amsfonts}
\usepackage{subfigure}
\usepackage{enumerate}
\usepackage{bm}
\usepackage{amsfonts}
\usepackage{psfrag}
\usepackage{algorithm}
\usepackage{algorithmic}
\usepackage{epsfig}\usepackage{cite}
\usepackage{array}
\usepackage{epsf}
\usepackage{epsfig}\usepackage{rotating}
\usepackage{longtable}
\usepackage{bm}
\usepackage{psfrag}
\usepackage{amsfonts}
\usepackage{subfigure}
\usepackage{afterpage}
\usepackage{rotating}
\usepackage{multirow}
\usepackage{enumerate}
\usepackage{tikz,ifthen}
\usepgflibrary{shapes.multipart}  
\usetikzlibrary{shapes.multipart} 

\def\ps@headings{\def\@oddhead{\mbox{}\scriptsize\rightmark \hfil \thepage}\def\@evenhead{\scriptsize\thepage \hfil \leftmark\mbox{}}\def\@oddfoot{}\def\@evenfoot{}}

\newtheorem{theorem}{Theorem}

\newtheorem{corollary}{Corollary}[theorem]

\long
\def\@makecaption#1#2{  \vskip\abovecaptionskip
\sbox\@tempboxa{#1: #2}
\ifdim \wd\@tempboxa >\hsize
#1: #2\par
\else
\global \@minipagefalse
\hb@xt@\hsize{\box\@tempboxa\hfil}  \fi
\vskip\belowcaptionskip}

\makeatother

\providecommand{\remarkname}{Remark}
\providecommand{\theoremname}{Theorem}

\begin{document}

\title{Fundamental Limits of Caching: Improved Bounds with Coded Prefetching}

\author{Jesús Gómez-Vilardebó}

\maketitle

\IEEEpeerreviewmaketitle{}
\begin{abstract}
We consider a cache network in which a single server is connected
to multiple users via a shared error free link. The server has access
to a database with $N$ files of equal length $F$, and serves $K$
users each with a cache memory of $MF$ bits. A novel centralized
coded caching scheme is proposed for scenarios with more users than
files $N\leq K$ and cache capacities satisfying $\frac{1}{K}\leq M\leq\frac{N}{K}$.
The proposed scheme outperforms the best rate-memory region known
in the literature if $N\leq K\leq\frac{N^{2}+1}{2}$.
\end{abstract}

\section{Introduction}

Content caching techniques are recently increasing attention to combat
peak hour traffic in content delivery services. The basic idea is
simple. If contents are made available at user terminals during low
traffic periods, then the peak rate can be reduced. However, content
requests are unknown to the server and thus content caching at user
memories must be carefully chosen in order to be useful regardless
of the contents requested during peak hours. The simplest caching
scheme consists of storing each file partially at each user memory.
Then, the server transmits the remaining requested data uncoded \cite{local1,local2}.
For single user caching systems, this strategy is optimal. However,
for multi-user systems, the seminal work in \cite{MaddhNiesen14}
by Maddah-Ali and Niesen shows that important gains can be obtained
by a new coded caching strategy. Specifically, there authors show
that, besides the local caching gain that is obtained by placing contents
at user caches before they are requested, it is possible to obtain
a global caching gain by creating broadcast opportunities. This is,
by carefully choosing the content caches at different users, and using
network coding techniques it is possible to transform the initial
multi-cast network, where every user is requesting a different file,
into a broadcast network, where every user requests exactly the same
``coded'' file, obtaining the new global caching gain.

The fundamental caching scheme developed in \cite{MaddhNiesen14}
was latter extended to more realistic situations, including a decentralized
scheme in \cite{MaddahDescentralized}, non-uniforms demands in \cite{NiesenNonUniform},
and online coded caching in \cite{PedarsaniOnline}. In addition,
new schemes pushing further the fundamental limits of caching systems
have appeared in \cite{DenizUncod,GastparCISS16,GhasemiISIT15,Gunduz_coded,Letaief,YuMA16,tianArxiv16}. 

The work here proposed investigates the fundamental achievable rate
for the particular situation where there are more users than files,
and the caching memories at users are small compared to the number
of files in the system. Besides its theoretical relevance, this situation
can be readily found in the real world. For instance, global content
delivery services such a Netflix serve a few multimedia contents to
millions of users across the world. In addition, it was shown in \cite{NiesenNonUniform}
that a near optimal caching strategy consists in dividing the files
into groups with similar popularity, and then applying the coded caching
strategy to each group separately. Since the amount of users in each
groups remains the same, when there are many groups, the cache size
dedicated to each group is small as well as the number of files per
user in each group. 

The rest of this paper is organized as follows. In Section \ref{sec:System-Model-and},
we present the system model together with the more relevant previous
results. In Section \ref{sec:Main-Result}, we summarize the main
results of this paper. Section \ref{sec:Propose-Caching-Scheme} describes
the caching scheme proposed, first, by providing a detailed example
and, then extending it to the general case. Finally, conclusions are
drawn in Section \ref{sec:Conclusions}.

\section{System Model and Previous Results\label{sec:System-Model-and}}

We consider a communication system with one server connected to $K$
users, denoted as $U_{1},...,U_{K}$, through a shared, error-free
link. There is a database at the server with $N$ files, each of length
$F$ bits, denoted as $W_{1},....,W_{N}$. Each users is equipped
with a local cache of capacity $MF$ bits and is assumed to request
only one full file. Here, we consider the special case where $M\in\left[0,\frac{N}{K}\right]$
and there are more users than files $N\leq K$. For convenience, we
define parameter $g=\frac{N}{MK}$.

We consider the communication model introduced in \cite{MaddhNiesen14}.
The caching system operates in two phases: the \emph{placement phase}
and the \emph{delivery phase}. In the placement phase, users have
access to the server database, and each user fills their cache. As
in \cite{Letaief}, we allow coding in the prefetching phase. Then,
each user $U_{k}$ requests a single full file $W_{\mathbf{d}(k)}$
where $\mathbf{d}=\left(\mathbf{d}(1),....,\mathbf{d}(K)\right)$
denotes the demand vector. We denote the number of distinct request
in $\mathbf{d}$ as $N_{e}(\mathbf{d})$. In the delivery phase, only
the server has access to the database. After being informed of the
user demands, the server transmits a signal $X$ of size $RF$ bits
over the shared link to satisfy all user requests simultaneously.
The signal $X$ is a function of the demand vector $\mathbf{d}$,
all the files in the data base, and the content in the user caches.
The rate $R$ is referred to as the rate of the shared link and is
a function of the demand vector $\mathbf{d}$. Using the local cache
content and the received signal $X$, each user $U_{k}$ reconstructs
its requested file $W_{\mathbf{d}(k)}$.

For a caching system ($M,N,K$), given a particular demand $\mathbf{d}$,
we define a communication rate $R$ is achievable if and only if there
exists a message $X$ of length $RF$ bits such that every user $U_{k}$
is able to reconstruct its desired file $W_{\mathbf{d}(k)}$. We denote
as $R(\mathbf{d})$ the achievable rate for a particular demand $\mathbf{d}$.
Then, the rate needed for the worst demand is given by $R^{\ast}=\max_{\mathbf{d\in}\mathcal{D}}R(\mathbf{d})$. 

For the caching system described, there is a trade-off between the
memory $M$ at users, and the worst demand rate $R^{\ast}$. Observe,
that if users have no caching capacity $M=0$, the server needs to
send the full requested files and thus, the worst demand rate is $R^{\ast}=N$.
Instead, if users can have a complete copy of the server's database
$M=N$, then no information needs to be transmitted from the server
$R^{\ast}=0$. We are interested in characterizing this rate memory
trade-off. To that end, we define the rate-memory pair $(R^{\ast},M)$
and the rate-memory function $R^{\ast}(M).$

\subsection{Previous Results}

For the special case considered here $M\in\left[0,\frac{N}{K}\right]$
and $N\leq K$, the best known rate-memory function in the literature
can be obtained by memory sharing between four achievable rate-memory
pairs: the trivial rate-memory pair $(N,0)$, the rate-memory pairs
obtained in \cite{Letaief} for $M=\frac{1}{K}$ 
\begin{equation}
(R_{\text{CFL}}^{\ast},M_{\text{CFL}})=\left(N-\frac{N}{K},\frac{1}{K}\right),\label{eq:CFL}
\end{equation}
the rate-memory pair obtained by the schemes proposed in \cite{DenizUncod}
and \cite{YuMA16} for $M=\frac{N}{K}$
\begin{equation}
(R_{\text{GBC}}^{\ast},M_{\text{GBC}})=\left(N-\frac{N(N+1)}{2K},\frac{N}{K}\right),\label{eq:GBC}
\end{equation}
and the rate-memory pairs obtained in \cite{tianArxiv16} 
\begin{equation}
(R_{\text{MDS}}^{\ast},M_{\text{MDS}})=\left(\frac{N\left(K-t\right)}{K},\frac{t\left[(N-1)t+K-N\right]}{K\left(K-1\right)}\right),\text{ }t=0,1,...,K.\label{eq:Tian}
\end{equation}
The lower convex envelope of all these rate-memory pairs, provides
the best rate-memory function in the literature. 

Our scheme is mainly motivated by the scheme described in \cite{Letaief},
which makes use of coded prefetching at users' cache. As shown in
\cite{Letaief}, the scheme achieving (\ref{eq:CFL}) is optimal for
$M=\frac{1}{K}.$ The schemes proposed in \cite{DenizUncod} and \cite{YuMA16}
assume uncoded prefetching. They are essentially the same at $M=\frac{N}{K}$.
The scheme proposed in \cite{YuMA16} was shown to be optimal among
all the uncoded prefetching schemes. Finally, the scheme developed
in \cite{tianArxiv16} makes use of non binary codes, in particular
distant separable (MDS) codes and rank metric codes to obtain the
rate-memory pairs in (\ref{eq:Tian}), which are shown to be optimal
at certain points. There have been other coded prefetching schemes
proposed in the literature, see \cite{Gunduz_coded} and \cite{GastparCISS16}
but either they do not improve the current best known rate-memory
trade-off or they apply to other situations. Despite all this efforts,
the optimal rate-memory trade-off for a caching systems remains an
open problem. There have been also efforts to obtain theoretical lower
bounds on the delivery rate. The cut-set bound was studied in \cite{MaddhNiesen14}.
A tighter lower bound was obtained in \cite{Tandon15}. Through a
computational approach a lower-bound for the special case $N=K=3$
is derived in \cite{Tian15}. Finally, in \cite{GhasemiISIT15} yet
another lower-bound is proposed.

\section{Main Result\label{sec:Main-Result}}

The following theorem presents the delivery rate obtained by the proposed
coded caching scheme for a particular demand $\mathbf{d}$.

\begin{theorem}\label{th:main}

For a caching problem with $K$ users and $N$ files, $K\geq N$,
local cache size of $M$ files at each user, and parameter $g=\frac{N}{MK}$.
Given a particular demand $\mathbf{d}$, the delivery rate

\begin{equation}
R=\frac{KN_{e}(\mathbf{d})\left(\begin{array}{c}
N-1\\
g-1
\end{array}\right)-g\left(\begin{array}{c}
N_{e}(\mathbf{d})+1\\
g+1
\end{array}\right)}{K\left(\begin{array}{c}
N-1\\
g-1
\end{array}\right)}\label{eq:thmain}
\end{equation}
is achievable for $g\in\left\{ 1,...,N\right\} $. Furthermore, for
$g\in[1,N]$ equals the lower convex envelope of its values at $g\in\left\{ 1,...,N\right\} $.

\end{theorem}

We prove this result and the next corollary in the following section
by describing the new caching scheme.

\begin{corollary}\label{co:rate_function}

For a caching problem with $K$ users and $N$ files, $K\geq N$,
local cache size of $M$ files at each user, and parameter $g=\frac{N}{MK}$,
the delivery rate-memory pairs 
\begin{equation}
\left(R^{\ast},M\right)=\left(N-\frac{N}{K}\frac{N+1}{g+1},\frac{N}{Kg}\right)\label{eq:rate-memory function}
\end{equation}
are achievable for $g\in\left\{ 1,...,N\right\} $. Furthermore, for
$g\in[1,N]$ the rate-memory pairs in the the lower convex envelope
of its values at $g\in\left\{ 1,...,N\right\} $ are achievable. \end{corollary} 
\begin{rem}
The rate-memory function in (\ref{eq:rate-memory function}) coincides
with (\ref{eq:CFL}) for $g=N$, and with (\ref{eq:GBC}) for $g=1$.
For $g=1$, $M=\frac{N}{K}$ our scheme is essentially the same as
the one described in \cite{YuMA16} and \cite{DenizUncod}. However,
the scheme proposed in \cite{Letaief} to achieve (\ref{eq:CFL})
differs slightly from the one considered here for $K>N$. Indeed,
while \cite{Letaief} divides each file into $NK$ subfiles, our scheme
requires only $K$ subfiles per file.
\end{rem}
The next three corollaries compare the proposed scheme with the scheme
presented in \cite{tianArxiv16}, the cut set lower bound derived
in \cite{MaddhNiesen14}, and the lower bound obtained in \cite{Tandon15}.
These results are proved in the Appendices.

\begin{corollary}\label{co:Tian}For a caching problem with $K$
users and $N$ files, $K\geq N$, and local cache size of $\frac{1}{K}\leq M\leq\frac{N}{K}$,
a sufficient condition for the proposed scheme to outperform the rate-memory
region obtained by the lower convex envelope of the rate-memory pairs
in (\ref{eq:Tian}), is 
\[
K\leq\frac{N^{2}+1}{2}.
\]
\end{corollary}

\begin{corollary}\label{co:CB}Let $R_{\text{CB}}(M)$ denote the
rate-memory function obtained for the cut set lower bound. The rate
difference between the cut set rate and the rate-memory pairs in Corollary
\ref{co:rate_function} is
\begin{eqnarray}
R_{\text{CB}}\left(M\right)-R^{\star}\left(M\right) & = & \frac{N}{K}\left(\frac{N}{g}-\frac{N+1}{g+1}\right)\nonumber \\
 & = & \frac{\left(\begin{array}{c}
N\\
g+1
\end{array}\right)}{K\left(\begin{array}{c}
N-1\\
g-1
\end{array}\right)}.\label{eq:cut-set}
\end{eqnarray}

\end{corollary}

Observe that, as first reported in \cite[Theorems 3 and 4]{Letaief},
for $g=N$, $M=\frac{1}{K}$ the cut set lower bound is achieved by
the proposed strategy. From the second equality (\ref{eq:cut-set}),
given that $K\left(\begin{array}{c}
N-1\\
g-1
\end{array}\right)$ is the number of subfile partitions in our scheme, we can see that
$\left(\begin{array}{c}
N\\
g+1
\end{array}\right)$ is the distance in number of subfile transmissions to the cut set
bound.

\begin{corollary}\label{co:Tandon}Let $R_{\text{STC}}(M)$ denote
the lower bound on the rate-memory function presented in \cite{Tandon15}.
For $K=N$ and $M=\frac{N}{(N-1)K}=\frac{1}{N-1}$, this lower bound
is achievable by the rate-memory function in Corollary \ref{co:rate_function}.
This is $R^{\star}\left(\frac{1}{N-1}\right)=R_{\text{STC}}\left(\frac{1}{N-1}\right)$
. 

\end{corollary}

\begin{figure}[t]
\psfrag{Mohammad}{hola}
\centering

\includegraphics{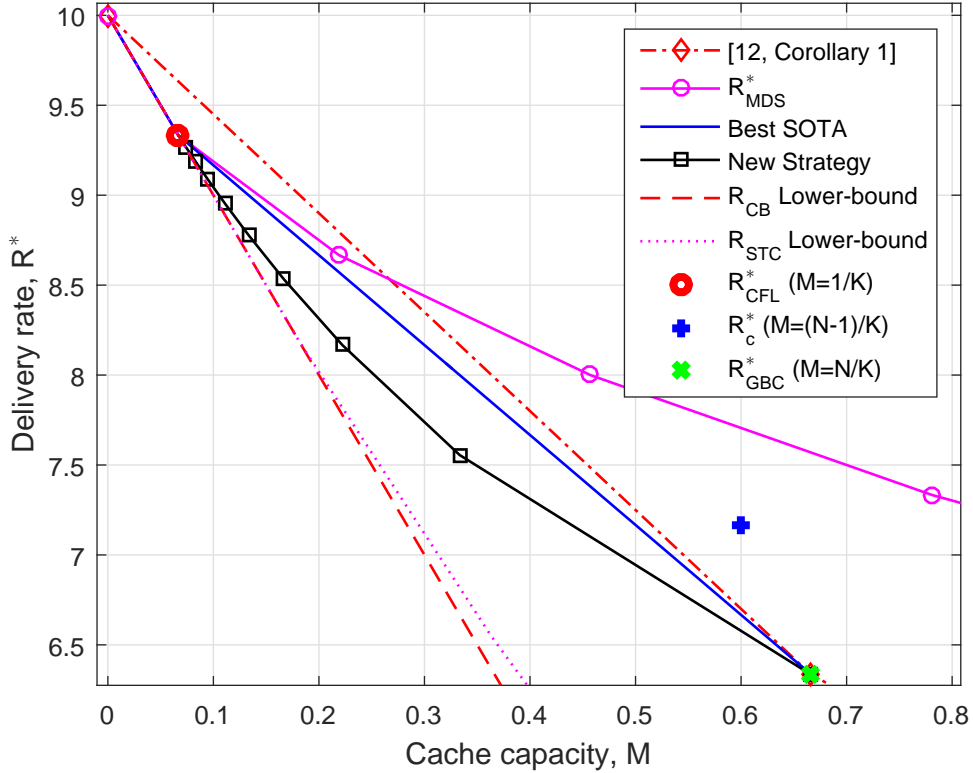}

\caption{Rate-memory functions of the proposed scheme (New Strategy) compared
with existing schemes and lower bounds in the literature for $N=10$
and $K=15$.\label{fig:ratememory}}
\end{figure}

We conclude this section by illustrating in Fig. \ref{fig:ratememory}
the rate-memory function for the proposed scheme and for the state
of the art (SOTA). We consider the case $N=10$ files and $K=15$
users. In addition to the best state of the art rate-memory function,
we provide the rate-memory function recently discovered in \cite{YuMA16},
the previously introduced rate-memory pairs $(R_{\text{MDS}}^{\ast},M_{\text{MDS}})$,
$(R_{\text{CFL}}^{\ast},M_{\text{CFL}})$, and $(R_{\text{GBC}}^{\ast},M_{\text{GBC}})$
as well as the rate-memory pair $\left(R_{\text{c}}^{\ast},\frac{N-1}{K}\right)$
found in \cite{Gunduz_coded} by using a coded prefetching scheme.
We also include in this figure for comparison, the cut set lower bound
and the information-theoretical lower bound obtained in \cite{Tandon15}.
We observe that, for the special situation considered here, the new
proposed scheme obtains a significant improvement with respect to
the previous best SOTA.

\section{Proposed Caching Scheme\label{sec:Propose-Caching-Scheme}}

We describe the proposed scheme, first with a particular example.
Then, we provide a detailed description of the general case.

\subsection{Example}

\begin{table*}\centering

\begin{tabular}{|c|c|c|c|}
\hline 
\textbf{Prefetching} &  & \textbf{Group} & \tabularnewline
\hline 
\textbf{User $U_{j}$} & \textbf{\{1,2\}} & \textbf{\{1,3\}} & \textbf{\{2,3\}}\tabularnewline
\hline 
\hline 
\textbf{}%
\begin{tabular}{c}
\textbf{Subfile}\tabularnewline
\textbf{assignment}\tabularnewline
\end{tabular}\textbf{ } & %
\begin{tabular}{c}
$W_{1,\{1,2\}}^{\left(j\right)}$\tabularnewline
$W_{2,\{1,2\}}^{\left(j\right)}$\tabularnewline
\end{tabular} & %
\begin{tabular}{c}
$W_{1,\{1,3\}}^{\left(j\right)}$\tabularnewline
$W_{3,\{1,3\}}^{\left(j\right)}$\tabularnewline
\end{tabular} & %
\begin{tabular}{c}
$W_{2,\{2,3\}}^{\left(j\right)}$\tabularnewline
$W_{3,\{2,3\}}^{\left(j\right)}$\tabularnewline
\end{tabular}\tabularnewline
\hline 
\textbf{Coded cached subfiles} & $W_{1,\{1,2\}}^{\left(j\right)}\oplus W_{2,\{1,2\}}^{\left(j\right)}$ & W$_{1,\{1,3\}}^{\left(j\right)}\oplus W_{3,\{1,3\}}^{\left(j\right)}$ & $W_{2,\{2,3\}}^{\left(j\right)}\oplus W_{3,\{2,3\}}^{\left(j\right)}$\tabularnewline
\hline 
\end{tabular}

\caption{Subfile assignement and prefetching at user $U_j$, for the proposed coded caching scheme when $K=6$, $N=3$ and $M=\frac{1}{4}$, i.e. $g=2$.}

\label{tbl:ex_prefeching}\end{table*}

Consider a caching system with $N=3$ files, $K=6$ users and a caching
capacity of $MF$ bits with $M=\frac{1}{4}$, which corresponds to
choosing \textbf{$g=2$.} For this particular case, the best known
coded caching scheme obtains the rate-memory pair $\left(\frac{28}{12}+\frac{1}{36},\frac{1}{4}\right)$
by memory sharing between the rate-memory pair $(R_{\text{MDS}}^{\ast},M_{\text{MDS}})=(R_{\text{CFL}}^{\ast},M_{\text{CFL}})=\left(\frac{5}{2},\frac{1}{6}\right)$
from (\ref{eq:Tian}) with $t=1$, and the rate-memory pair $(R_{\text{MDS}}^{\ast},M_{\text{MDS}})=\left(2,\frac{7}{15}\right)$
from (\ref{eq:Tian}) with $t=2$. Here, we show that the rate-memory
pair $\left(\frac{28}{12},\frac{1}{4}\right)$ is achievable. The
best known outer bound \cite{Tandon15} obtains the rate-memory pair
$\left(\frac{27}{12},\frac{1}{4}\right).$

The prefetching strategy is illustrated in Table \ref{tbl:ex_prefeching}.
We break each file $W_{f}$, $f=\left\{ 1,2,3\right\} $ into $K\left(\begin{array}{c}
N-1\\
g-1
\end{array}\right)=6\times2=12$ subfiles of equal size. Next, two different subfiles for each file
are assigned to every user. Each subfile of file $W_{f}$ assigned
to user $U_{i}$ is further associated to one of the two sets of $g=2$
subfiles $\mathcal{A}\in\left\{ \left\{ 1,2\right\} ,\left\{ 1,3\right\} ,\left\{ 2,3\right\} \right\} $
which contain file index $f$, this is $f\in\mathcal{A}$, and denoted
as $W_{f,\mathcal{A}}^{\left(i\right)}.$ Finally, each user $U_{i}$,
$i=\left\{ 1,...,6\right\} $ computes the binary sum of the $g$
subfiles assigned to each of the $3$ possible sets $\mathcal{A}$,
and stores the resultant coded subfile. In the following, these stored
coded subfiles are referred to as\emph{ coded cached subfiles}.

Given the above prefetching scheme, we illustrate our proposed delivery
strategy for a representative demand scenario, where users $U_{1}$
and $U_{2}$ request file $W_{1}$, users $U_{3}$ and $U_{4}$ request
file $W_{2}$, and users $U_{5}$ and $U_{6}$ request file $W_{3}$.
We show later in this section, by addressing the general case, that
this indeed corresponds to the worst possible demand. We represent
this demand by the demand vector $\mathbf{d}=[1,1,2,2,3,3]$. Then
$W_{\mathbf{d}(i)}$ returns the file requested by user $U_{i}$. 

First, observe that, users all together, store all subfiles XORed
at some coded cached subfile, and that every subfile is only XORed
at one of the coded cached subfiles. Having this in mind, the delivery
scheme proposed divides the coded cached subfiles into three types
and obtains the subfiles XORed at each of the coded cached subfiles
types separately. 

\uline{Coded cached subfiles Type I:} First, consider those coded
cached subfiles XORing a subfile requested by some user together with
subfiles that are not requested by any user. For the specific demand
considered in this example, there are non of these coded cached subfiles,
as every file is requested by some user. If there would be some, the
server would simply broadcast, one by one, the subfiles requested
by users XORed in every coded cached subfile Type I. 

\begin{table*}\centering

\begin{tabular}{|c|c|c|c|c|c|c|c|}
\hline 
\textbf{User} & \textbf{Request} & \textbf{}%
\begin{tabular}{c}
\textbf{Coded cached}\tabularnewline
\textbf{subfiles}\tabularnewline
\end{tabular} & \multicolumn{3}{c|}{\textbf{Phase 1}} & \multicolumn{2}{c|}{\textbf{Phase 2}}\tabularnewline
\hline 
 &  &  & \textbf{Broadcast} & \textbf{Compute} & \textbf{Broadcasted} & \textbf{Broadcast} & \textbf{Compute}\tabularnewline
\hline 
$U_{1}^{*}$ & $W_{1}$ & %
\begin{tabular}{c}
$W_{1,\{1,2\}}^{\left(1\right)}\oplus W_{2,\{1,2\}}^{\left(1\right)}$\tabularnewline
$W_{1,\{1,3\}}^{\left(1\right)}\oplus W_{3,\{1,3\}}^{\left(1\right)}$\tabularnewline
\end{tabular} & %
\begin{tabular}{c}
$W_{2,\{1,2\}}^{\left(1\right)}$\tabularnewline
$W_{3,\{1,3\}}^{\left(1\right)}$\tabularnewline
\end{tabular} & %
\begin{tabular}{c}
$\underline{W_{1,\{1,2\}}^{\left(1\right)}}$\tabularnewline
$\underline{W_{1,\{1,3\}}^{\left(1\right)}}$\tabularnewline
\end{tabular} & \multirow{2}{*}[10pt]{%
\begin{tabular}{c}
$W_{1,\{1,2\}}^{\left(3\right)}$\tabularnewline
$W_{1,\{1,2\}}^{\left(4\right)}$\tabularnewline
$W_{1,\{1,3\}}^{\left(5\right)}$\tabularnewline
$W_{1,\{1,3\}}^{\left(6\right)}$\tabularnewline
\end{tabular}} & \multirow{2}{*}{%
\begin{tabular}{c}
$\underline{W_{1,\{1,2\}}^{\left(1\right)}}\oplus W_{1,\{1,2\}}^{\left(2\right)}$\tabularnewline
$\underline{W_{1,\{1,3\}}^{\left(1\right)}}\oplus W_{1,\{1,3\}}^{\left(3\right)}$\tabularnewline
\end{tabular}} & %
\begin{tabular}{c}
$W_{1,\{1,2\}}^{\left(2\right)}$\tabularnewline
$W_{1,\{1,3\}}^{\left(2\right)}$\tabularnewline
\end{tabular}\tabularnewline
\cline{1-5} \cline{8-8} 
$U_{2}$ & $W_{1}$ & %
\begin{tabular}{c}
$W_{1,\{1,2\}}^{\left(2\right)}\oplus W_{2,\{1,2\}}^{\left(2\right)}$\tabularnewline
$W_{1,\{1,3\}}^{\left(2\right)}\oplus W_{3,\{1,3\}}^{\left(2\right)}$\tabularnewline
\end{tabular} & %
\begin{tabular}{c}
$W_{2,\{1,2\}}^{\left(2\right)}$\tabularnewline
$W_{3,\{1,3\}}^{\left(2\right)}$\tabularnewline
\end{tabular} & %
\begin{tabular}{c}
$W_{1,\{1,2\}}^{\left(2\right)}$\tabularnewline
$W_{1,\{1,3\}}^{\left(2\right)}$\tabularnewline
\end{tabular} &  &  & %
\begin{tabular}{c}
$\underline{W_{1,\{1,2\}}^{\left(1\right)}}$\tabularnewline
$\underline{W_{1,\{1,3\}}^{\left(1\right)}}$\tabularnewline
\end{tabular}\tabularnewline
\hline 
$U_{3}^{\ast}$ & $W_{2}$ & %
\begin{tabular}{c}
$W_{1,\{1,2\}}^{\left(3\right)}\oplus W_{2,\{1,2\}}^{\left(3\right)}$\tabularnewline
$W_{2,\{2,3\}}^{\left(3\right)}\oplus W_{3,\{2,3\}}^{\left(3\right)}$\tabularnewline
\end{tabular} & %
\begin{tabular}{c}
$W_{1,\{1,2\}}^{\left(3\right)}$\tabularnewline
$W_{3,\{2,3\}}^{\left(3\right)}$\tabularnewline
\end{tabular} & %
\begin{tabular}{c}
$\underline{W_{2,\{1,2\}}^{\left(3\right)}}$\tabularnewline
$\underline{W_{2,\{2,3\}}^{\left(3\right)}}$\tabularnewline
\end{tabular} & \multirow{2}{*}[10pt]{%
\begin{tabular}{c}
$W_{2,\{1,2\}}^{\left(1\right)}$\tabularnewline
$W_{1,\{1,2\}}^{\left(2\right)}$\tabularnewline
$W_{2,\{2,3\}}^{\left(5\right)}$\tabularnewline
$W_{2,\{2,3\}}^{\left(6\right)}$\tabularnewline
\end{tabular}} & \multirow{2}{*}{%
\begin{tabular}{c}
$\underline{W_{2,\{1,2\}}^{\left(3\right)}}\oplus W_{2,\{1,2\}}^{\left(4\right)}$\tabularnewline
$\underline{W_{2,\{2,3\}}^{\left(3\right)}}\oplus W_{2,\{2,3\}}^{\left(4\right)}$\tabularnewline
\end{tabular}} & %
\begin{tabular}{c}
$W_{2,\{1,2\}}^{\left(4\right)}$\tabularnewline
$W_{2,\{2,3\}}^{\left(4\right)}$\tabularnewline
\end{tabular}\tabularnewline
\cline{1-5} \cline{8-8} 
$U_{4}$ & $W_{2}$ & %
\begin{tabular}{c}
$W_{1,\{1,2\}}^{\left(4\right)}\oplus W_{2,\{1,2\}}^{\left(4\right)}$\tabularnewline
$W_{2,\{2,3\}}^{\left(4\right)}\oplus W_{3,\{2,3\}}^{\left(4\right)}$\tabularnewline
\end{tabular} & %
\begin{tabular}{c}
$W_{1,\{1,2\}}^{\left(4\right)}$\tabularnewline
$W_{3,\{2,3\}}^{\left(4\right)}$\tabularnewline
\end{tabular} & %
\begin{tabular}{c}
$W_{2,\{1,2\}}^{\left(4\right)}$\tabularnewline
$W_{2,\{2,3\}}^{\left(4\right)}$\tabularnewline
\end{tabular} &  &  & %
\begin{tabular}{c}
$\underline{W_{2,\{1,2\}}^{\left(3\right)}}$\tabularnewline
$\underline{W_{2,\{2,3\}}^{\left(3\right)}}$\tabularnewline
\end{tabular}\tabularnewline
\hline 
$U_{5}^{\ast}$ & $W_{3}$ & %
\begin{tabular}{c}
$W_{1,\{1,3\}}^{\left(5\right)}\oplus W_{3,\{1,3\}}^{\left(5\right)}$\tabularnewline
$W_{2,\{2,3\}}^{\left(5\right)}\oplus W_{3,\{2,3\}}^{\left(5\right)}$\tabularnewline
\end{tabular} & %
\begin{tabular}{c}
$W_{1,\{1,3\}}^{\left(5\right)}$\tabularnewline
$W_{2,\{2,3\}}^{\left(5\right)}$\tabularnewline
\end{tabular} & %
\begin{tabular}{c}
$\underline{W_{3,\{1,3\}}^{\left(5\right)}}$\tabularnewline
$\underline{W_{3,\{2,3\}}^{\left(5\right)}}$\tabularnewline
\end{tabular} & \multirow{2}{*}[10pt]{%
\begin{tabular}{c}
$W_{3,\{1,3\}}^{\left(1\right)}$\tabularnewline
$W_{3,\{1,3\}}^{\left(2\right)}$\tabularnewline
$W_{3,\{2,3\}}^{\left(3\right)}$\tabularnewline
$W_{3,\{2,3\}}^{\left(4\right)}$\tabularnewline
\end{tabular}} & \multirow{2}{*}{%
\begin{tabular}{c}
$\underline{W_{3,\{1,3\}}^{\left(5\right)}}\oplus W_{3,\{1,3\}}^{\left(6\right)}$\tabularnewline
$\underline{W_{3,\{2,3\}}^{\left(5\right)}}\oplus W_{3,\{2,3\}}^{\left(6\right)}$\tabularnewline
\end{tabular}} & %
\begin{tabular}{c}
$W_{3,\{1,3\}}^{\left(6\right)}$\tabularnewline
$W_{3,\{2,3\}}^{\left(6\right)}$\tabularnewline
\end{tabular}\tabularnewline
\cline{1-5} \cline{8-8} 
$U_{6}$ & $W_{3}$ & %
\begin{tabular}{c}
$W_{1,\{1,3\}}^{\left(6\right)}\oplus W_{3,\{1,3\}}^{\left(6\right)}$\tabularnewline
$W_{2,\{2,3\}}^{\left(6\right)}\oplus W_{3,\{2,3\}}^{\left(6\right)}$\tabularnewline
\end{tabular} & %
\begin{tabular}{c}
$W_{1,\{1,3\}}^{\left(6\right)}$\tabularnewline
$W_{2,\{2,3\}}^{\left(6\right)}$\tabularnewline
\end{tabular} & %
\begin{tabular}{c}
$W_{3,\{1,3\}}^{\left(6\right)}$\tabularnewline
$W_{3,\{2,3\}}^{\left(6\right)}$\tabularnewline
\end{tabular} &  &  & %
\begin{tabular}{c}
$\underline{W_{3,\{1,3\}}^{\left(5\right)}}$\tabularnewline
$\underline{W_{3,\{2,3\}}^{\left(5\right)}}$\tabularnewline
\end{tabular}\tabularnewline
\hline 
\end{tabular}

\caption{Delivery scheme of requested subfiles XORed at coded cached subfiles Type II.}

\label{tbl:typeII;example}

\end{table*}

\uline{Coded cached subfiles Type II:} Second, consider those coded
cached subfiles XORing a subfile requested by the user which stores
them. For our specific example, there are two of such coded cached
subfiles at each user. The delivery strategy for all the subfiles
XORed at these coded cached subfiles is represented in Table \ref{tbl:typeII;example}.
The delivery scheme is divided into two consecutive phases, and each
phase is further divided into a broadcast and computation task:

\textbf{Phase 1:} For each of the coded cached subfiles Type II, the
server first broadcasts (forth column), one by one, all the subfiles
XORed at each coded cached subfile except for the one requested by
the user in which the coded cached subfile is stored. Users obtain
directly from this broadcasted transmission, all their requested subfiles
which are XORed at the coded cached subfiles stored at other users
with a different demand (sixth column). Using these broadcasted subfiles
(forth column) and their coded cached subfiles (third column), each
user computes the requested subfiles (fifth column) which are XORed
at each of the coded cached subfiles. 

\textbf{Phase 2:} Observe that, after Phase 1, every set of users
with the same demand has computed a different and not overlapping
subset of requested subfiles. In order to all the users with a common
demand share their computed subfiles in Phase 1, the server arbitrarily
selects one user \emph{leader} for each subset of users with the same
requested file. Let $\text{\ensuremath{\mathcal{U}}}=\left\{ 1,3,5\right\} $
denote the set of user leaders, then for every user $i\notin\mathcal{U}$
and every subset $\mathcal{A}$ of $g$ files including $W_{\mathbf{d}(i)}$,
the server transmits (seventh column) the binary sum
\[
\underline{W_{\mathbf{d}(u),\mathcal{A}}^{(u)}}\oplus W_{\mathbf{d}(i),\mathcal{A}}^{(i)}
\]
where $u\in\mathcal{U}$ is the user leader with the same request
$\mathbf{d}(u)=\mathbf{d}(i)$. For clarity, we underline the subfiles
obtained in Phase 1 at user leaders. Finally (eight column), given
that each of the broadcasted coded subfiles XOR a subfile already
available at the user leader together with a subfile already available
at a non-user leader in the same demand group, user leaders can compute
the remaining requested subfiles. Every other user, would first compute
the subfile belonging to the user leader and then use it to obtain
the rest of the requested subfiles in their demand group.

\begin{table*}\centering

\begin{tabular}{|c|c|c|c|c|c|c|c|}
\hline 
\textbf{User} & \textbf{}%
\begin{tabular}{c}
\textbf{Coded cached}\tabularnewline
\textbf{subfiles}\tabularnewline
\end{tabular}\textbf{ } & \multicolumn{2}{c|}{\textbf{Phase 1}} & \multicolumn{4}{c|}{\textbf{Phase 2}}\tabularnewline
\hline 
 &  & \textbf{Broadcast} & \textbf{Compute} & \textbf{Broadcast } & \textbf{Compute} & \textbf{}%
\begin{tabular}{c}
\textbf{Reordering }\tabularnewline
\textbf{Phase 1\textbackslash{}Broadcast}\tabularnewline
\end{tabular} & \textbf{Compute}\tabularnewline
\hline 
$U_{1}^{*}$ & $\underline{W_{2,\{2,3\}}^{\left(1\right)}}\oplus W_{3,\{2,3\}}^{\left(1\right)}$ & $W_{3,\{2,3\}}^{\left(1\right)}\oplus\underline{W_{3,\{1,3\}}^{\left(3\right)}}$ & \multirow{2}{*}{%
\begin{tabular}{c}
$\underline{W_{2,\{2,3\}}^{\left(1\right)}}$\tabularnewline
$\oplus$\tabularnewline
$\underline{W_{3,\{1,3\}}^{\left(3\right)}}$\tabularnewline
\end{tabular}} & \multirow{6}{*}{\textbf{}%
\begin{tabular}{c}
$\underline{W_{1,\{1,2\}}^{\left(5\right)}}$\tabularnewline
$\oplus$\tabularnewline
$\underline{W_{2,\{2,3\}}^{\left(1\right)}}$\tabularnewline
$\oplus$\tabularnewline
$\underline{W_{3,\{1,3\}}^{\left(3\right)}}$\tabularnewline
\end{tabular}} & \multirow{2}{*}{$\underline{W_{1,\{1,2\}}^{\left(5\right)}}$} & \multirow{2}{*}[2pt]{%
\begin{tabular}{c}
$W_{1,\{1,3\}}^{\left(1\right)}\oplus\underline{W_{1,\{1,2\}}^{\left(5\right)}}$\tabularnewline
$W_{1,\{1,3\}}^{\left(4\right)}\oplus\underline{W_{1,\{1,2\}}^{\left(5\right)}}$\tabularnewline
$W_{1,\{1,2\}}^{\left(6\right)}\oplus\underline{W_{1,\{1,2\}}^{\left(5\right)}}$\tabularnewline
\end{tabular}} & \multirow{2}{*}{%
\begin{tabular}{c}
$W_{1,\{1,3\}}^{\left(1\right)}$\tabularnewline
$W_{1,\{1,3\}}^{\left(4\right)}$\tabularnewline
$W_{1,\{1,2\}}^{\left(6\right)}$\tabularnewline
\end{tabular}}\tabularnewline
\cline{1-3} 
$U_{2}$ & $W_{2,\{2,3\}}^{\left(2\right)}\oplus W_{3,\{2,3\}}^{\left(2\right)}$ & %
\begin{tabular}{c}
$W_{2,\{2,3\}}^{\left(2\right)}\oplus\underline{W_{2,\{2,3\}}^{\left(1\right)}}$\tabularnewline
$W_{3,\{2,3\}}^{\left(2\right)}\oplus$$\underline{W_{3,\{1,3\}}^{\left(3\right)}}$\tabularnewline
\end{tabular} &  &  &  &  & \tabularnewline
\cline{1-4} \cline{6-8} 
$U_{3}^{\ast}$ & $W_{1,\{1,3\}}^{\left(3\right)}\oplus\underline{W_{3,\{1,3\}}^{\left(3\right)}}$ & $W_{1,\{1,3\}}^{\left(1\right)}\oplus\underline{W_{1,\{1,2\}}^{\left(5\right)}}$ & \multirow{2}{*}{%
\begin{tabular}{c}
$\underline{W_{1,\{1,2\}}^{\left(5\right)}}$\tabularnewline
$\oplus$\tabularnewline
$\underline{W_{3,\{1,3\}}^{\left(3\right)}}$\tabularnewline
\end{tabular}} &  & \multirow{2}{*}{$\underline{W_{2,\{2,3\}}^{\left(1\right)}}$} & \multirow{2}{*}[2pt]{%
\begin{tabular}{c}
$W_{2,\{1,2\}}^{\left(5\right)}\oplus\underline{W_{2,\{2,3\}}^{\left(1\right)}}$\tabularnewline
$W_{2,\{2,3\}}^{\left(2\right)}\oplus\underline{W_{2,\{2,3\}}^{\left(1\right)}}$\tabularnewline
$W_{2,\{1,2\}}^{\left(5\right)}\oplus\underline{W_{2,\{2,3\}}^{\left(1\right)}}$\tabularnewline
\end{tabular}} & \multirow{2}{*}{%
\begin{tabular}{c}
$W_{2,\{1,2\}}^{\left(5\right)}$\tabularnewline
$W_{2,\{2,3\}}^{\left(2\right)}$\tabularnewline
$W_{2,\{1,2\}}^{\left(5\right)}$\tabularnewline
\end{tabular}}\tabularnewline
\cline{1-3} 
$U_{4}$ & $W_{1,\{1,3\}}^{\left(4\right)}\oplus W_{3,\{1,3\}}^{\left(4\right)}$ & %
\begin{tabular}{c}
$W_{1,\{1,3\}}^{\left(4\right)}\oplus\underline{W_{1,\{1,2\}}^{\left(5\right)}}$\tabularnewline
$W_{3,\{1,3\}}^{\left(4\right)}\oplus$$\underline{W_{3,\{1,3\}}^{\left(3\right)}}$\tabularnewline
\end{tabular} &  &  &  &  & \tabularnewline
\cline{1-4} \cline{6-8} 
$U_{5}^{\ast}$ & $\underline{W_{1,\{1,2\}}^{\left(5\right)}}\oplus W_{2,\{1,2\}}^{\left(5\right)}$ & $W_{2,\{2,3\}}^{\left(1\right)}\oplus\underline{W_{2,\{1,2\}}^{\left(5\right)}}$ & \multirow{2}{*}{%
\begin{tabular}{c}
$\underline{W_{1,\{1,2\}}^{\left(5\right)}}$\tabularnewline
$\oplus$\tabularnewline
$\underline{W_{2,\{1,2\}}^{\left(1\right)}}$\tabularnewline
\end{tabular}} &  & \multirow{2}{*}{$\underline{W_{3,\{1,3\}}^{\left(3\right)}}$} & \multirow{2}{*}{%
\begin{tabular}{c}
$W_{3,\{2,3\}}^{\left(1\right)}\oplus\underline{W_{3,\{1,3\}}^{\left(3\right)}}$\tabularnewline
$W_{3,\{2,3\}}^{\left(2\right)}\oplus$$\underline{W_{3,\{1,3\}}^{\left(3\right)}}$\tabularnewline
\multirow{1}{*}[2pt]{$W_{3,\{1,3\}}^{\left(4\right)}\oplus$$\underline{W_{3,\{1,3\}}^{\left(3\right)}}$}\tabularnewline
\end{tabular}} & \multirow{2}{*}{%
\begin{tabular}{c}
$W_{3,\{2,3\}}^{\left(1\right)}$\tabularnewline
$W_{3,\{2,3\}}^{\left(2\right)}$\tabularnewline
$W_{3,\{1,3\}}^{\left(4\right)}$\tabularnewline
\end{tabular}}\tabularnewline
\cline{1-3} 
$U_{6}$ & $W_{1,\{1,2\}}^{\left(6\right)}\oplus W_{2,\{1,2\}}^{\left(6\right)}$ & %
\begin{tabular}{c}
$W_{2,\{1,2\}}^{\left(6\right)}\oplus\underline{W_{2,\{2,3\}}^{\left(1\right)}}$\tabularnewline
$W_{1,\{1,2\}}^{\left(6\right)}\oplus\underline{W_{1,\{1,2\}}^{\left(5\right)}}$\tabularnewline
\end{tabular} &  &  &  &  & \tabularnewline
\hline 
\end{tabular}

\caption{Delivery scheme of requested subfiles XORed at coded cached subfiles Type III.}

\label{tbl:typeIII;example}

\end{table*}

\uline{Coded cached subfiles Type III:} Finally, consider those
coded cached subfiles which XOR together subfiles not requested by
the user storing them. The delivery of the subfiles in coded cached
subfiles Type III is illustrated in Table \ref{tbl:typeIII;example}.
In our specific example, there is only one of such coded cached subfiles
at each user cache (third column). Again, the delivery scheme is partitioned
into two consecutive phases:

\textbf{Phase 1:} Select a set of user \emph{leaders} $\text{\ensuremath{\mathcal{U}}}=\left\{ 1,3,5\right\} $,
for simplicity we consider the one used for coded cached subfiles
Type II. Next, for every subset $\mathcal{V\subseteq\mathcal{U}}$
of $g+1=3$ user leaders (in our particular example there is only
one of such subsets $\mathcal{V}=\left\{ 1,2,3\right\} $), there
is an associated set of file requests $\mathcal{B}(\mathcal{V})=\{W_{1},W_{2},W_{3}\}$.
The servers arbitrarily selects one subfile, (the underlined subfiles
in the second column), each from a different file in $\mathcal{B}(\mathcal{V})$,
from the subfiles XORed at each of the coded cached subfiles Type
III of the user leaders in $\mathcal{V}$. Next, for every other subfile
XORed at coded cached subfiles Type III of all users (leaders or not),
the server broadcasts the binary sum of such subfile and the one selected
(underlined) that belongs to the same file (see third column). Finally,
using these coded broadcasted subfiles, together with their own coded
cached subfiles Type III, each user computes the binary sum of all
the selected (underlined) subfiles belonging to files distinct from
their requested file (forth column).

\textbf{Phase 2:} Next, the server broadcasts the binary sum of all
the selected/underlined subfiles (fifth column). Using this coded
broadcasted subfile together with the coded subfile computed in Phase
1 (third column) each user computes the selected/underlined subfile
belonging to its requested file (sixth column). Finally, each user
obtains their remaining requested subfiles (eight column) by computing
the binary sum of the recently computed subfile (sixth column) and
each of the coded broadcasted subfiles in Phase 1 that XOR two subfiles
belonging to its requested file (seventh column).

\subsection{General Scheme}

Next, we generalize the proposed centralized coded caching scheme.
Consider a scenario with $K$ users, $N$ files of size $F$ bits
and a cache capacity at users of $M=\frac{N}{gK}$. To achieve the
rate $R$ stated in Theorem 1, we present a prefetching and delivery
scheme for $g\in\left\{ 0,1,...,N\right\} $, since for general $\frac{1}{K}\leq M\leq\frac{N}{K}$,
the minimum rate can be achieved by memory sharing.

\emph{Prefetching scheme}: In \cite{Letaief}, the prefetching scheme
stores at each user a different coded subfile XORing together $N$
subfiles, one from each of the $N$ files available in the system.
The prefetching scheme proposed here extends this idea by storing
at each user a different coded subfile XORing together $g\in\left\{ 0,1,...,N\right\} $
subfiles each from a different file. Given that there is not just
one, but $\left(\begin{array}{c}
N\\
g
\end{array}\right)$ different combinations of $g$ different files, and no particular
combination shall be preferred, we store one coded subfile for each
possible combinations of $g$ files. Specifically, we partition each
file $W_{f}$, $f\in\left\{ 1,...,N\right\} $ into $K\left(\begin{array}{c}
N-1\\
g-1
\end{array}\right)$ non-overlapping subfiles of equal size and assign $\left(\begin{array}{c}
N-1\\
g-1
\end{array}\right)$ subfiles of each file to each user $U_{i}$, $i\in\left\{ 1,...,K\right\} $.
Next, consider each possible subset $\mathcal{A}\subseteq\left\{ 1,...,N\right\} $
of $g$ different files, i.e. $\left|\mathcal{A}\right|=g$. Each
of the $\left(\begin{array}{c}
N-1\\
g-1
\end{array}\right)$ subfiles of file $W_{f}$ assigned to user $U_{i}$ is also assigned
to a different subset $\mathcal{A}$ satisfying $f\in\mathcal{A}$.
We identify this subfile as $W_{f,\mathcal{A}}^{(i)}$. This is possible,
since there are $\left(\begin{array}{c}
N-1\\
g-1
\end{array}\right)$ subsets $\mathcal{A}$ which include a particular file $W_{f}$,
i.e. satisfying $f\in\mathcal{A}$. Finally, at user $U_{i}$, for
every subset of files $\mathcal{A}$, we compute the binary sum of
the $g$ subfiles assigned to subset $\mathcal{A}$, and store the
coded cached subfile

\[
Z_{\mathcal{A}}^{(i)}=\bigoplus_{f\in\mathcal{A}}W_{f,\mathcal{A}}^{(i)}.
\]
Finally, because each user stores $\left(\begin{array}{c}
N\\
g
\end{array}\right)$ coded cached subfiles, one for each subset $\mathcal{A}$, and each
subfile has $\frac{F}{K\left(\begin{array}{c}
N-1\\
g-1
\end{array}\right)}$ bits, the required cache load at each user equals $MF=\frac{\left(\begin{array}{c}
N\\
g
\end{array}\right)}{K\left(\begin{array}{c}
N-1\\
g-1
\end{array}\right)}F=\frac{N}{gK}F$ bits, which satisfies the memory constraint imposed.

\emph{Delivery scheme}: As pointed out in the example, all subfiles
can be found only once at some code cached subfile. The delivery scheme
proposed divides the coded cached subfiles into three types and obtains
the requested subfiles XORed at each of the coded cached subfile types
separately.

\uline{Coded cached subfiles Type I}: First, consider those requested
subfiles which are XORed at coded cached subfiles that also XOR subfiles
not requested by any user. The server simply broadcast, one by one,
all the requested subfiles in every coded cached subfile Type I. Suppose
only $N_{e}(\mathbf{d})$ distinct files out of the total $N$ files
are requested by all users. From the placement phase, we know that
XORed at all the coded cached subfiles, there are $\left(\begin{array}{c}
N-1\\
g-1
\end{array}\right)$ subfiles of each file at every user, from which there are $\left(\begin{array}{c}
N_{e}(\mathbf{d})-1\\
g-1
\end{array}\right)$ requested subfiles XORed to other requested subfiles. Thus, there
are $\left(\begin{array}{c}
N-1\\
g-1
\end{array}\right)-\left(\begin{array}{c}
N_{e}(\mathbf{d})-1\\
g-1
\end{array}\right)$ distinct subfiles of every file at every user which are encoded together
with some not requested subfile. Finally, since there are $N_{e}(\mathbf{d})$
different requests and $K$ users, the total number of subfile transmissions
needed is
\[
T_{\text{I}}=KN_{e}(\mathbf{d})\left(\left(\begin{array}{c}
N-1\\
g-1
\end{array}\right)-\left(\begin{array}{c}
N_{e}(\mathbf{d})-1\\
g-1
\end{array}\right)\right).
\]

\uline{Coded cached subfiles Type II:} Second, consider those coded
cached subfiles which XOR a subfile requested by the user which stores
them. Notice that, all requested subfiles still missing are encoded
at some coded cache subfile together with $g-1$ other also requested
subfiles.The delivery strategy for all the subfiles XORed at the coded
cached subfiles Type II is divided into two consecutive phases. In
the first phase, each user $U_{i}$ obtains the subfiles $W_{\mathbf{d}(i),\mathcal{A}}^{(i)}$
of their own demand $\mathbf{d}(i)$ encoded at coded cached subfiles
$Z_{\mathcal{A}}^{(i)}$ in their own cache. In the second phase,
every two distinct users $U_{i}$, $U_{j}$ requesting the same file
$\mathbf{d}(i)=\mathbf{d}(j)$ share the subfiles obtained in Phase
1.

\textbf{Phase 1:} For each of the coded cached subfiles Type II, the
server first broadcasts, one by one, all the subfiles XORed at each
coded cached subfile except the one requested by the user in which
the coded cached subfile is stored. This is, the server broadcast
each of the $g-1$ subfiles XORed to $W_{\mathbf{d}(i),\mathcal{A}}^{(i)}$
in $Z_{\mathcal{A}}^{(i)}$. Given that there are $\left(\begin{array}{c}
N_{e}(\mathbf{d})-1\\
g-1
\end{array}\right)$ of such subfiles at each user, we require 
\begin{equation}
K(g-1)\left(\begin{array}{c}
N_{e}(\mathbf{d})-1\\
g-1
\end{array}\right)\label{eq:TypeIIPhase1}
\end{equation}
subfile transmissions. Next, by computing the binary sum of $Z_{\mathcal{A}}^{(i)}$
and the broadcasted subfiles XORed in the coded cached subfile $Z_{\mathcal{A}}^{(i)}$,
each user obtains the requested subfile $W_{\mathbf{d}(i),\mathcal{A}}^{(i)}$.
At the end of this phase, user $U_{i}$ has obtained the $\left(\begin{array}{c}
N_{e}(\mathbf{d})-1\\
g-1
\end{array}\right)$ requested subfiles which where cached at some coded cached subfile
in its own cache. In addition, users obtain, directly, from the broadcasted
subfiles, all their requested subfiles which are XORed at the coded
cached subfiles stored at users with a different demand. Specifically,
let $\mathcal{K}(f)$ denote the set of users requesting file $W_{f}$,
user $U_{i}$ obtains $\left(\begin{array}{c}
N_{e}(\mathbf{d})-2\\
g-2
\end{array}\right)\left(K-\left|\mathcal{K}(f)\right|\right)$ subfiles of his own requested file $W_{\mathbf{d}(i)}$ directly
from the broadcast transmission. For consistency check, observe that
the total number of subfiles obtained directly from the broadcast
transmissions by all users coincides with the total number of subfiles
transmitted in Phase 1.
\begin{eqnarray}
\left(\begin{array}{c}
N_{e}(\mathbf{d})-2\\
g-2
\end{array}\right)\sum_{\forall f\in\mathbf{d}}\left(K-\left|K(f)\right|\right) & = & \left(\begin{array}{c}
N_{e}(\mathbf{d})-2\\
g-2
\end{array}\right)\left(N_{e}(\mathbf{d})K-K\right)\nonumber \\
 & = & (\ref{eq:TypeIIPhase1})\label{eq:TypeIIPhase1_2}
\end{eqnarray}
where (\ref{eq:TypeIIPhase1_2}) follows from $\left(\begin{array}{c}
n\\
m
\end{array}\right)=\left(\begin{array}{c}
n-1\\
m-1
\end{array}\right)\frac{n}{m}$.

\textbf{Phase 2:} After Phase 1, because all subfiles XORed at coded
cached subfiles of different users are distinct, users with the same
request have obtained a different subset of $\left(\begin{array}{c}
N_{e}(\mathbf{d})-1\\
g-1
\end{array}\right)$ subfiles of their common requested file. Consider the set of users
$\mathcal{K}(f)$ requesting file $W_{f}$. In order to all users
$U_{i}$, requesting file $W_{f}$, $i\in\mathcal{K}(f)$ share the
subfiles obtained Phase 1, the server arbitrarily selects one user
\emph{leader} $u\in$ $\mathcal{K}(f)$ per file $W_{f}$. For every
other user $U_{i},$ $i\in K(f)/u$ and every subset $\mathcal{A}$
of $g$ requested files including $W_{\mathbf{d}(i)}$, the server
transmits the binary sum
\begin{equation}
Y_{f,\mathcal{A}}^{(i,u)}=W_{f,\mathcal{A}}^{(i)}\oplus W_{f,\mathcal{A}}^{(u)}.\label{eq:PhaseIItx}
\end{equation}
Given that there are $\left(\begin{array}{c}
N_{e}(\mathbf{d})-1\\
g-1
\end{array}\right)$ subsets $\mathcal{A}$ of $g$ requested files, we require a total
of
\begin{equation}
\left(\left|K(f)\right|-1\right)\left(\begin{array}{c}
N_{e}(\mathbf{d})-1\\
g-1
\end{array}\right)\label{eq:T_II_1}
\end{equation}
coded subfile transmissions per file.

Next, for every requested file $W_{f}$, the user leader $u\in\mathcal{K}(f)$
computes the remaining requested subfiles by computing the binary
sum of the subfile $W_{f,\mathcal{A}}^{(u)}$ obtained in Phase 1
with each of the broadcasted coded subfiles 
\begin{equation}
W_{f,\mathcal{A}}^{(i)}=Y_{f,\mathcal{A}}^{(i,u)}\oplus W_{f,\mathcal{A}}^{(u)}\label{eq:PhaseII_3}
\end{equation}
 for all $i\in\mathcal{K}(f)/u$ and all sets of $g$ requested files
$\mathcal{A}$, such that $\mathbf{d}(u)\in\mathcal{A}.$ Every other
user $U_{j}$, $j\neq u$, requesting file $W_{f}$, $j\in\mathcal{K}(f)$
first obtains the subfile of the user leader $W_{f,\mathcal{A}}^{(u)}$
by computing
\[
W_{f,\mathcal{A}}^{(u)}=Y_{f,\mathcal{A}}^{(j,u)}\oplus W_{f,\mathcal{A}}^{(j)},
\]
and then uses $W_{f,\mathcal{A}}^{(u)}$ to obtain the rest of the
subfiles by computing (\ref{eq:PhaseII_3}) for all $i\in\mathcal{K}(f)/u/j$
and all sets of $g$ requested files $\mathcal{A}$, such that $f\in\mathcal{A}.$
Because there are $N_{e}(\mathbf{d})$ distinct requested files, the
total number of coded subfile transmissions (\ref{eq:PhaseIItx})
is 

\begin{equation}
\sum_{\forall f\in\mathbf{d}}\left(\left|\mathcal{K}(f)\right|-1\right)\left(\begin{array}{c}
N_{e}(\mathbf{d})-1\\
g-1
\end{array}\right)=\left(K-N_{e}(\mathbf{d})\right)\left(\begin{array}{c}
N_{e}(\mathbf{d})-1\\
g-1
\end{array}\right).\label{eq:T_II_2}
\end{equation}
Finally, adding together (\ref{eq:TypeIIPhase1}) and (\ref{eq:T_II_2}),
the total number of subfile length transmissions needed to obtain
the requested subfiles XORed in the coded cached subfiles Type II
is 
\[
T_{\text{II}}=\left(Kg-N_{e}(\mathbf{d})\right)\left(\begin{array}{c}
N_{e}(\mathbf{d})-1\\
g-1
\end{array}\right).
\]

\uline{Coded cached subfiles Type III:} Finally, consider those
coded cached subfiles which XOR together subfiles not requested by
the user storing them. Specifically, consider any subset $\mathcal{V}$
of $g+1$ users with different demands. Let $\mathcal{B}(\mathcal{V})$
denote the set of $g+1$ demands of the users in $\mathcal{V}$. For
any user, $U_{i}$, $i\in\mathcal{V}$ there is exactly one coded
cached subfile
\[
Z_{\mathcal{B}(\mathcal{V})\backslash\mathbf{d}(i)}^{(i)}=\bigoplus_{f\in\mathcal{B}(\mathcal{V})\backslash\mathbf{d}(i)}W_{f,\mathcal{B}(\mathcal{V})\backslash\mathbf{d}(i)}^{(i)}
\]
which does not XOR a subfile of $\mathbf{d}(i)$, but XORs together
one subfile requested by each of the other $g-1$ users in $\mathcal{V}$.
In this phase, all user obtain such subfiles. Observe that for coded
cached subfiles Type III to exist, there might be at least $g+1$
users with different demands, this is $N_{e}(\mathbf{d})\geq g+1$.
Again, the delivery scheme is partitioned into two consecutive phases:

\textbf{Phase 1:} First, the server selects a set of $N_{e}(\mathbf{d})$
user \emph{leaders} $\text{\ensuremath{\mathcal{U}}}$ each with a
different demand. Next, for every subset $\mathcal{V\subseteq\mathcal{U}}$
of $g+1$ user leaders, the server arbitrarily associates to each
user, $U_{j}$, $j\in\mathcal{V}$ a different file $\mathbf{r}_{\mathcal{V}}(j)\in\mathcal{B}(\mathcal{V})$,
such that $\mathbf{r}_{\mathcal{V}}(j)\neq\mathbf{d}(j)$ and $\mathbf{r}_{\mathcal{V}}(i)\neq\mathbf{\mathbf{r}_{\mathcal{V}}}(j)$
for all $i\neq j$, $i\in\mathcal{V}$. Since $\left|\mathcal{B}(\mathcal{V})\right|=g+1$,
this is always possible. The server then broadcasts the coded subfiles

\[
Y_{\mathcal{V}}^{(i,j)}=W_{\mathbf{\mathbf{r}_{\mathcal{V}}}(j),\mathcal{B}(\mathcal{V})\backslash\mathbf{d}(i)}^{(i)}\oplus W_{\mathbf{\mathbf{\mathbf{r}_{\mathcal{V}}}}(j),\mathcal{B}(\mathcal{V})\backslash\mathbf{d}(j)}^{(j)}
\]
 for all $j\in\mathcal{V\backslash}i$ with $\mathbf{\mathbf{r}_{\mathcal{V}}}(j)\neq\mathbf{d}(i)$,
$\mathbf{\mathbf{\mathbf{r}_{\mathcal{V}}}}(j)\neq\mathbf{\mathbf{\mathbf{r}_{\mathcal{V}}}}(i)$
and all $i\in\mathcal{V}$. Because there are $\left(\begin{array}{c}
N_{e}(\mathbf{d})\\
g+1
\end{array}\right)$ subsets $\mathcal{V}\subseteq\mathcal{U}$ of $g+1$ users, this
results in a total of $\left(\begin{array}{c}
N_{e}(\mathbf{d})\\
g+1
\end{array}\right)(g-1)(g+1)$ subfile length transmissions. Next, for every other user, $U_{i}$,
$i\notin\mathcal{U}$ satisfying $\mathbf{d}(i)\in\mathcal{\mathcal{B}\left(V\right)}$,
the server broadcasts
\[
Y_{\mathcal{V}}^{(i,j)}=W_{\mathbf{\mathbf{r}_{\mathcal{V}}}(j),\mathcal{B}(\mathcal{V})\backslash\mathbf{d}(i)}^{(i)}\oplus W_{\mathbf{\mathbf{\mathbf{r}_{\mathcal{V}}}}(j),\mathcal{B}(\mathcal{V})\backslash\mathbf{d}(j)}^{(j)}
\]
for all $j\in\mathcal{V}$ with $\mathbf{r_{\mathcal{V}}}(j)\neq\mathbf{d}(i)$.
Because there are $K-N_{e}(\mathbf{d})$ not users leaders $i\notin\mathcal{U}$,
and there are $\left(\begin{array}{c}
N_{e}(\mathbf{d})-1\\
g
\end{array}\right)$ subsets $\mathcal{V}\subseteq\mathcal{U}$ of $g+1$ users satisfying
$\mathbf{d}(j)\in\mathcal{V}$, this results in a total of $g\left(K-N_{e}(\mathbf{d})\right)\left(\begin{array}{c}
N_{e}(\mathbf{d})-1\\
g
\end{array}\right)$ coded subfile transmissions.

Next, for all subsets of $g+1$ users $\mathcal{V}\subseteq\mathcal{U}$,
every user $U_{i}$, satisfying $\mathbf{d}(i)\in\mathcal{B}(\mathcal{V})$
computes 
\begin{eqnarray*}
Z_{\mathcal{B}\left(\mathcal{V}\right)\backslash\mathbf{d}\left(i\right)}^{(i)}\bigoplus_{j\in\mathcal{V}\backslash i,\mathbf{r}_{\mathcal{V}}\left(j\right)\neq\mathbf{d}(i)}Y_{\mathcal{V}}^{(i,j)} & = & \bigoplus_{f\in\mathcal{B}\left(\mathcal{V}\right)\backslash\mathbf{d}\left(i\right)}W_{f,\mathcal{B}\left(V\right)\backslash\mathbf{d}\left(i\right)}^{(i)}\bigoplus_{j\in\mathcal{V}\backslash i,\mathbf{r}_{\mathcal{V}}\left(j\right)\neq\mathbf{d}(i)}Y_{\mathcal{V}}^{(i,j)}\\
 & = & \bigoplus_{j\in\mathcal{V},\mathbf{r}_{\mathcal{V}}\left(j\right)\neq\mathbf{d}(i)}W_{\mathbf{r}_{\mathcal{V}}\left(j\right),\mathcal{B}\left(V\right)\backslash\mathbf{d}\left(i\right)}^{(i)}\bigoplus_{j\in\mathcal{V}\backslash i,\mathbf{r}_{\mathcal{V}}\left(j\right)\neq\mathbf{d}(i)}Y_{\mathcal{V}}^{(i,j)}\\
 & = & W_{\mathbf{r}_{\mathcal{V}}\left(i\right),\mathcal{B}\left(V\right)\backslash\mathbf{d}\left(i\right)}^{(i)}\bigoplus_{j\in\mathcal{V}\backslash i,\mathbf{r}_{\mathcal{V}}\left(j\right)\neq\mathbf{d}(i)}\left(W_{\mathbf{r}_{\mathcal{V}}\left(j\right),\mathcal{B}\left(V\right)\backslash\mathbf{d}\left(i\right)}^{(i)}\oplus Y_{\mathcal{V}}^{(i,j)}\right)\\
 & = & W_{\mathbf{r}_{\mathcal{V}}\left(i\right),\mathcal{B}\left(V\right)\backslash\mathbf{d}\left(i\right)}^{(i)}\bigoplus_{j\in\mathcal{V}\backslash i,\mathbf{r}_{\mathcal{V}}\left(j\right)\neq\mathbf{d}(i)}W_{\mathbf{\mathbf{\mathbf{r}_{\mathcal{V}}}}(j),\mathcal{B}(\mathcal{V})\backslash\mathbf{d}(j)}^{(j)}\\
 & = & \bigoplus_{j\in\mathcal{V},\mathbf{r}_{\mathcal{V}}\left(j\right)\neq\mathbf{d}(i)}W_{\mathbf{\mathbf{\mathbf{r}_{\mathcal{V}}}}(j),\mathcal{B}(\mathcal{V})\backslash\mathbf{d}(j)}^{(j)}
\end{eqnarray*}
where for every $i\notin\mathcal{V}$, we defined $\mathbf{r}_{\mathcal{V}}\left(i\right)=\mathbf{r}_{\mathcal{V}}\left(j\right)$
where $j\in\mathcal{V}$ and $\mathbf{d}(i)=\mathbf{d}(j)$. 

\textbf{Phase 2:} Finally, for each subset of $g+1$ users leaders
$\mathcal{V}\subseteq\mathcal{U}$, the server broadcasts the coded
subfile 
\[
Y_{\mathcal{V}}=\bigoplus_{j\in\mathcal{V}}W_{\mathbf{\mathbf{r}_{\mathcal{V}}}(j),\mathcal{B}(\mathcal{V})\backslash\mathbf{d}(j)}^{(j)}.
\]
This results in a total of $\left(\begin{array}{c}
N_{e}(\mathbf{d})\\
g+1
\end{array}\right)$ additional coded subfiles transmissions. Now, for all subsets of
$g+1$ users $\mathcal{V}\subseteq\mathcal{U}$, every user $U_{i}$,
satisfying $\mathbf{d}(i)\in\mathcal{B}(\mathcal{V})$ obtains the
subfile $W_{\mathbf{r_{\mathcal{V}}}(j)=\mathbf{d}(i),\mathcal{B}(\mathcal{V})\backslash\mathbf{d}(j)}^{(j)}$
which is cached at user leader, $U_{j}$ $j\in\mathcal{V}$ satisfying
$\mathbf{r_{\mathcal{V}}}(j)=\mathbf{d}(i),$ by computing 
\begin{eqnarray*}
\bigoplus_{j\in\mathcal{V},\mathbf{r}_{\mathcal{V}}\left(j\right)\neq\mathbf{d}(i)}W_{\mathbf{\mathbf{\mathbf{r}_{\mathcal{V}}}}(j),\mathcal{B}(\mathcal{V})\backslash\mathbf{d}(j)}^{(j)}\oplus Y_{\mathcal{V}} & = & W_{\mathbf{r_{\mathcal{V}}}(j)=\mathbf{d}(i),\mathcal{B}(\mathcal{V})\backslash\mathbf{d}(j)}^{(j)}\\
 & = & W_{\mathbf{d}(i),\mathcal{B}(\mathcal{V})\backslash\mathbf{d}(j)}^{(j)}.
\end{eqnarray*}
User $U_{i}$ obtains the rest of his requested subfiles $W_{\mathbf{d}(i),\mathcal{B}(\mathcal{V})\backslash\mathbf{d}(l)}^{(l)}$
XORed at the coded cached subfile Type III of user $U_{l}$, $Z_{\mathcal{B}(\mathcal{V})\backslash\mathbf{d}(l)}^{(l)}$
by choosing $j\in\mathcal{V}$ satisfying $\mathbf{r_{\mathcal{V}}}(j)=\mathbf{d}(i)$
and computing 
\begin{eqnarray*}
W_{\mathbf{d}(i),\mathcal{B}(\mathcal{V})\backslash\mathbf{d}(j)}^{(j)}\oplus Y_{\mathcal{V}}^{(l,j)} & = & W_{\mathbf{d}(i),\mathcal{B}(\mathcal{V})\backslash\mathbf{d}(j)}^{(j)}\oplus W_{\mathbf{\mathbf{r}_{\mathcal{V}}}(j),\mathcal{B}(\mathcal{V})\backslash\mathbf{d}(l)}^{(l)}\oplus W_{\mathbf{\mathbf{\mathbf{r}_{\mathcal{V}}}}(j),\mathcal{B}(\mathcal{V})\backslash\mathbf{d}(j)}^{(j)}\\
 & = & W_{\mathbf{d}(i),\mathcal{B}(\mathcal{V})\backslash\mathbf{d}(l)}^{(l)}
\end{eqnarray*}
for all subsets of $g+1$ users $\mathcal{V}\subseteq\mathcal{U}$
and all users $l\neq i$ satisfying $\mathbf{d}(i)=\mathbf{d}(l)$. 

The total number of transmissions in both phases is
\begin{eqnarray}
T_{\text{III}} & = & \left(\begin{array}{c}
N_{e}(\mathbf{d})\\
g+1
\end{array}\right)\left((g-1)(g+1)+1\right)+\left(K-N_{0}\right)g\left(\begin{array}{c}
N_{e}(\mathbf{d})-1\\
g
\end{array}\right)\nonumber \\
 & = & \left(\begin{array}{c}
N_{e}(\mathbf{d})-1\\
g
\end{array}\right)g\left(K-\frac{N_{e}(\mathbf{d})}{g+1}\right)\label{eq:III_1}\\
 & = & \left(\begin{array}{c}
N_{e}(\mathbf{d})-1\\
g-1
\end{array}\right)\left(N_{e}(\mathbf{d})-g\right)\left(K-\frac{N_{e}(\mathbf{d})}{g+1}\right)\label{eq:III_2}
\end{eqnarray}
where (\ref{eq:III_1}) follows from $\left(\begin{array}{c}
n\\
k
\end{array}\right)=\left(\begin{array}{c}
n-1\\
k-1
\end{array}\right)\frac{n}{k}$ and (\ref{eq:III_2}) follows from $\left(\begin{array}{c}
n\\
k
\end{array}\right)=\left(\begin{array}{c}
n\\
k-1
\end{array}\right)\frac{n+1-k}{k}$. 

Finally, adding together all the subfile length transmissions required
for the three coded cached subfile types, we obtain
\begin{eqnarray}
T & = & KN_{e}(\mathbf{d})\left(\begin{array}{c}
N-1\\
g-1
\end{array}\right)-N_{e}(\mathbf{d})\left(\frac{N_{e}(\mathbf{d})+1}{g+1}\right)\left(\begin{array}{c}
N_{e}(\mathbf{d})-1\\
g-1
\end{array}\right)\nonumber \\
 & = & KN_{e}(\mathbf{d})\left(\begin{array}{c}
N-1\\
g-1
\end{array}\right)-g\left(\begin{array}{c}
N_{e}(\mathbf{d})+1\\
g+1
\end{array}\right)\label{eq:T}
\end{eqnarray}
which leads to the rate (\ref{eq:thmain}) stated in Theorem \ref{th:main}.
we obtain the rate-memory function stated in Corollary \ref{co:rate_function}.

Finally, to prove Corollary \ref{co:rate_function}, we need to show
that (\ref{eq:T}) increases monotonically as a function of $N_{e}(\mathbf{d})$.
For the easy of notation, let us substitute $N_{e}(\mathbf{d})$ by
$x$, and denote (\ref{eq:T}) as a function of $N_{e}(\mathbf{d})$
as $T(x)$. Then, observe that

\begin{eqnarray}
T(x) & = & Kx\left(\begin{array}{c}
N-1\\
g-1
\end{array}\right)-g\left(\begin{array}{c}
x+1\\
g+1
\end{array}\right)\nonumber \\
 & = & Kx\left(\begin{array}{c}
N-1\\
g-1
\end{array}\right)-g\left(\begin{array}{c}
x+2\\
g+1
\end{array}\right)\left(1-\frac{g+1}{x+2}\right)\label{eq:T(x)_1}\\
 & = & Kx\left(\begin{array}{c}
N-1\\
g-1
\end{array}\right)-g\left(\begin{array}{c}
x+2\\
g+1
\end{array}\right)+g\frac{g+1}{x+2}\left(\begin{array}{c}
x+2\\
g+1
\end{array}\right)\nonumber \\
 & = & Kx\left(\begin{array}{c}
N-1\\
g-1
\end{array}\right)-g\left(\begin{array}{c}
x+2\\
g+1
\end{array}\right)+g\left(\begin{array}{c}
x+1\\
g
\end{array}\right)\label{eq:T(x)_2}\\
 & = & T(x+1)+(x+1)\left(\begin{array}{c}
x\\
g-1
\end{array}\right)-K\left(\begin{array}{c}
N-1\\
g-1
\end{array}\right)\nonumber 
\end{eqnarray}
where (\ref{eq:T(x)_1}) follows from $\left(\begin{array}{c}
n\\
k
\end{array}\right)=\left(\begin{array}{c}
n+1\\
k
\end{array}\right)\left(1-\frac{k}{n+1}\right)$ and (\ref{eq:T(x)_2}) follows by applying $\left(\begin{array}{c}
n\\
k
\end{array}\right)=\left(\begin{array}{c}
n-1\\
k-1
\end{array}\right)\frac{n}{k}$ twice. Thus, for $x=1$ to $x=N-1$, we have 
\begin{eqnarray*}
T(x+1) & = & T(x)+K\left(\begin{array}{c}
N-1\\
g-1
\end{array}\right)-(x+1)\left(\begin{array}{c}
x\\
g-1
\end{array}\right)\\
 & \geq & T(x)
\end{eqnarray*}
where the last inequality follows since $\left(\begin{array}{c}
x\\
g-1
\end{array}\right)$ increases monotonically with $x$, and $x\leq N-1\leq K-1$. Finally,
particularizing (\ref{eq:T}) to $N_{e}(\mathbf{d})=N$, we obtain
the rate-memory function in \ref{eq:rate-memory function}.

\section{Conclusions\label{sec:Conclusions}}

In this work, we proposed a novel centralized coded caching scheme
for the case where there are more users than files and users are equipped
with small memories. The scheme uses coded prefetching and outperforms
previously proposed schemes for moderate-high number of users, $N\leq K\leq\frac{N^{2}+1}{2}$.
Our current future work includes the extension of the proposed coded
prefetching technique to larger memories and scenarios with more users
than files, as well as, the extension of the proposed centralized
technique to the decentralized setting.

\begin{appendices}\label{appendices}

\section{Proof of Corollary \ref{co:Tian}}

To prove this results, we first observe that by isolating parameter
$t$ in the MDS rate points, $t=\frac{K}{N}\left(N-R_{\text{MDS}}^{\ast}\right)$
and substituting it into the $M_{\text{MDS}}$ memory points, we can
obtain the following memory-rate function 
\begin{equation}
M_{\text{MDS}}^{(\text{lb})}\left(R\right)=\frac{\left(N-R\right)\left((N-R)\left(NK-K\right)+N(K-N)\right)}{N^{2}\left(K-1\right)}.\label{eq:MDS_lb}
\end{equation}
Due to the convexity of $M_{\text{MDS}}^{(\text{lb})}\left(R\right)$
function with respect to $R$, we have that $M_{\text{MDS}}^{(\text{lb})}\left(R\right)$
represents a lower-bound on the MDS memory-rate region that can be
obtained by the lower convex envelope of the memory-rate pairs (\ref{eq:Tian}). 

Particularizing (\ref{eq:MDS_lb}) to the rate points in (\ref{eq:rate-memory function}),
we can write $M_{\text{MDS}}^{(\text{lb})}$ as a function of $g$
as

\[
M_{\text{MDS}}^{(\text{lb})}(g)=\frac{N+1}{g+1}\frac{1}{K(K-1)}\left(\frac{N^{2}-1}{g+1}+K-N\right)
\]
for $g=1,...,N$. Thus, a sufficient conditions for the new rate-memory
pairs in (\ref{eq:rate-memory function}) to improve the MDS rate-memory
region is
\[
\frac{N}{Mg}\leq M_{\text{MDS}}^{(\text{lb})}(g)
\]
which leads us to 
\[
K\leq\frac{N^{2}g+1}{g+1}.
\]
Finally, since $\frac{N^{2}g+1}{g+1}$ increases monotonically with
$g$, we have that a sufficient condition for our strategy to improve
the MDS strategy in the complete region considered $M\in\left[0,\frac{N}{K}\right]$
and $N\leq K$, is
\[
K\leq\frac{N^{2}+1}{2}.
\]

\section{Proof of Corollary \ref{co:CB}}

The proof of Corollary \ref{co:CB} follows the lines of \cite{Letaief}.
From \cite{MaddhNiesen14}, the cut set lower bound is given by
\[
R_{\text{CB}}(M)=\max_{s\in\left\{ 1,...,\min(N,K)\right\} }s-\frac{s}{\left\lfloor \frac{N}{s}\right\rfloor }M.
\]
Particularized to the case $N\leq K$, we have
\begin{eqnarray*}
R_{\text{CB}}(M) & \geq & \max\left(1-\frac{1}{\left\lfloor \frac{N}{s}\right\rfloor }M,....,N-NM\right)\\
 & \geq & N-NM\\
 & = & R^{\ast}\left(M\right)-\frac{N}{K}\left(\frac{N}{g}-\frac{N+1}{g+1}\right)\\
 & = & R^{\ast}(M)-\frac{\left(\begin{array}{c}
N\\
g+1
\end{array}\right)}{K\left(\begin{array}{c}
N-1\\
g-1
\end{array}\right)}
\end{eqnarray*}
where $R^{\text{\ensuremath{\ast}}}\left(M\right)$ is the rate-memory
tradeoff function presented in Corollary \ref{co:rate_function} for
$g\in\left\{ 1,...,N\right\} $.

\section{Proof of Corollary \ref{co:Tandon}}

The lower bound obtained in \cite[Theorem 1]{Tandon15} is given by
\begin{equation}
R_{\text{STC}}(M)=\max_{\begin{array}{c}
s\in\left\{ 1,...,K)\right\} \\
l\in\left\{ 1,...,\left\lceil \frac{N}{s}\right\rceil \right\} 
\end{array}}\frac{1}{l}\left(N-sM-\frac{\mu\left(N-ls\right)^{+}}{s+\mu}-\left(N-Kl\right)^{+}\right)\label{eq:tandon}
\end{equation}
where $\mu=\min\left(\frac{N-ls}{l},K-s\right)$ $\forall s,l$. Observe
that for $l=1$ and $s=N-1$, $K\geq N$, we have $\mu=\min(N-s,K-s)=N-s=1$,
and the objective in (\ref{eq:tandon}) reads $N-\left(N-1\right)M-\frac{1}{N}$.
Thus, 
\begin{eqnarray*}
R_{\text{STC}}(M) & \geq & N-\left(N-1\right)M-\frac{1}{N}
\end{eqnarray*}
which particularized to $K=N$, $M=\frac{N}{\left(N-1\right)K}$,
$g=N-1$ leads us to 
\begin{eqnarray*}
R_{\text{STC}}\left(\frac{N}{(N-1)K}\right) & \geq & N-1-\frac{1}{N}\\
 & = & R^{\ast}\left(\frac{1}{N-1}\right).
\end{eqnarray*}

\end{appendices}

\bibliographystyle{IEEEtran}
\bibliography{IEEEabrv,C:/Users/cttc/Desktop/jgomez_cloud/LINES/WORKINGpapers/CACHING}

\end{document}